\documentclass[useAMS,usenatbib]{mn2e}

\usepackage{graphicx}
\usepackage{myaasmacros}

\title{Terrestrial Planet Formation in the Inclined Systems:
Application to OGLE-2006-BLG-109L System}

\author[Sheng  Jin and Jianghui Ji]{Sheng  Jin$^{1,2}$ and Jianghui  Ji
$^{1}$\thanks{E-mail: jijh@pmo.ac.cn}\\
$^{1}$ Purple Mountain Observatory, Chinese Academy of Sciences,
Nanjing 210008, China\\
$^{2}$ Graduate School of Chinese Academy of Sciences, Beijing
100049, China}

\begin{document}

\date{Accepted, Received £»in original form 2011 July 8}

\pagerange{\pageref{firstpage}--\pageref{lastpage}} \pubyear{2011}

\maketitle

 \label{firstpage}

\begin{abstract}
In this work, we extensively investigate the terrestrial planetary
formation for the inclined planetary systems (considering the
OGLE-2006-BLG-109L system as prototype) in the late stage. In the
simulations, we show that the occurrence of terrestrial planets is
quite common, in the final assembly stage. Moreover, we find that
40\% of the runs finally occupy one planet in the habitable zone
(HZ). On the other hand, the numerical results also indicate that
the inner region of the planetesimal disk, ranging from $\sim 0.1$
to $0.3$ AU, plays an important role in building up terrestrial
planets. By examining all simulations, we note that the survivals
are located either between 0.1$\sim$1.0 AU or beyond 7 AU, or at the
1:1 mean motion resonance of OGLE-2006-BLG-109Lb at $\sim$2.20 AU.
The outcomes suggest that it may exist moderate possibility for the
inclined systems to harbor terrestrial planets, even planets in the
HZs.

\end{abstract}


\section{Introduction}

To date, over 500 exoplanets have been discovered
(http://exoplanet.eu), revealing a diversity of planetary systems.
One fascinating fact that has revealed so far is the population of
"Hot Jupiters" -- gas giants moving in very small orbits (periods
$<$ 8 days) about their parent stars, of which the prototype was the
first exoplanet discovered, 51 Peg \citep{may95}. Recently,
improvement of measurement precisions for both Doppler and transit
discovery techniques have made the detection of more than a dozen
exoplanets in the mass range from $\sim$ 2 to 15 $M_{\oplus}$
(super-Earths), among which there are also some short-period hot
super-Earths -- Gl 876 d \citep{rive05}, Gl 581 e \citep{may09},
CoRoT-7b \citep{leger09} and the Kepler-11 system \citep{liss11},
etc. These exciting observations greatly motivate us to explore the
open issues, to understand the formation and evolution of planetary
system.

According to core-accretion model, it is generally believed that
planetary formation could be divided into several stages. The
starting planetary formation scenario is the agglomeration of $\mu$m
dust grains to produce kilometer-sized bodies called planetesimals
\citep{Saf69,wether80}, however, the buildup of particles from cm to
meter size is still unclear \citep{liss93,youd02}. After they
emerge, numerous planetesimals can further form 1000-kilometer-sized
planetary embryos (protoplanet) and even the cores of gas-giants via
direct collision-merger process. In this stage, a few larger bodies
would grow rapidly considering their dominant role of gravitational
focusing in the runaway growth \citep{Saf69,weth89}. At the time
that they are separated into various feeding zones, the largest
bodies become massive enough to stir up the nearby small objects,
and subsequently the oligarchic growth takes over. In the oligarchic
stage, the largest bodies accrete surrounding masses in the disk
much slower than they do in runaway growth, but still grow more
quickly than those minor bodies \citep{Kok98,gold04}. However, the
oligarchic growth may cease once these massive bodies accrete to
have their isolation masses, where $M_{iso}\propto
M_{\ast}^{-1/2}\Sigma_{p}^{3/2}a^{3}$ \citep{armi07}, and $M_{\ast}$
is the mass of the central star, $\Sigma_{p}$, the surface density,
and $a$, the semi-major axis. This leads to the formation of larger
solid cores for jovian planets at distant orbits. Once they reach
critical mass ($\sim 10 M_{\oplus}$), the solid cores begin to
accrete gas very fast, and then to grow into giant planets
\citep{Kok02,Ida04}. All aforementioned scenarios are now considered
to be completed within $5-10$ Myr, which is critical for the buildup
of gas-giants, as inferred lifetimes of gaseous disks
\citep{hais01,wyatt08}.

The following stage is featured by terrestrial planet formation
under the interference of gas-giants, which is considered to be the
longest evolution history with collisional coagulation. In this late
stage, the formed gas-giants, accompanied with a large number of
planetesimals and planetary embryos, have ceased migration since
most of gas in the disk disappears. These low-mass objects, under
the perturbation of jovian planets as well as their mutual
gravitational interaction, may experience a turbulent evolution
process that may last for hundreds of million years. During this
stage, the embryos are greatly stirred to have crossing orbits,
which bring about numerous giant impacts \citep{cham98}. Finally,
small bodies are left over after the former stages have been cleaned
up and several terrestrial planets emerge
\citep{cham01,raym04,zha09} in the system.

Under some circumstances, the formed planets could be evolved into a
highly-inclined configuration, rather than a planar disk
perpendicular to the stellar spin axis, which seems quite natural
since both the star and planets obtain their angular momenta from
the disk. One mechanism that could increase the planetary
inclination is the scenario of planet-planet scattering.
\citet{chatt08} show that after strong planet-planet gravitational
scattering, the planetary inclination, in the initial range from
$0^{\circ} \sim 4^{\circ}$, could be dramatically excited. In their
simulations, more than half of the final configurations have a
relative inclination exceeding $20^{\circ}$ and nearly a quarter
above $40^{\circ}$. Considering the tidal damping timescale for
inclination is usually much larger than the star's age
\citep{winn05}, such highly-inclined planets could be found in the
planetary systems which have undergone severe gravitational
scattering in their lifetimes.

An alternate scenario for the origin of highly-inclined planets is
that they exist in binary systems. The presence of a companion star
can affect the orbits of planets and the stability of the planetary
system. In tight binaries, a close encounter between a planet and
the companion can eject the planet out of the system
\citep{holman99}. While in wider binaries, secular effects like
Kozai resonance become more important. The Kozai mechanism was first
introduced to analyze how the orbits of inclined asteroids were
influenced by Jupiter \citep{kozai62}. Now it is widely accepted to
explain the formation mechanism of a relative highly-inclined
configuration, such as a Jupiter-size exoplanet in wide inclined
binaries \citep{Inna97}, etc. The Kozai mechanism is only notable
when the inclination between the orbits of the planet and the
companion, $i_{0}$, is greater than a critical inclination $i_{c}$
of around $40^{\circ}$ (depending on the system) \citep{thomas96}.
If this is the case, the inclination of the planet will oscillate,
between $i_{c}$ and $i_{0}$, and is coupled to eccentricity $e$
through the Kozai constant \citep{thomas96}
\begin{equation}
H_{K}=\sqrt{a(1-e^{2})}\cos i
\end{equation}
\citet{takeda08} showed that the interplay between secular
perturbations from a stellar companion and the gravitational
coupling planets could trigger the chaotic growth of mutual
inclination angles between initially coplanar planets.

Observations have revealed that some planetary systems have large
obliquities. Recently, direct measurements of the planets $\upsilon$
Andromedae ($\upsilon$ And) c and d unveil a $30^{\circ}$ mutual
inclination angle between them \citep{mcarthur10}. \citet{barnes10}
showed that planet-planet scattering is a plausible mechanism to
explain the observed orbits of $\upsilon$ And c and d, though it is
unsure whether the scattering was caused by instabilities among the
planets or by perturbations from a nearby low-mass star $\upsilon$
And B. In recent years, transit observations have measured
$\lambda\sin i$ for a couple of systems, where $\lambda$ is the
angle between the stellar rotation axis and the orbital angular
momentum of a transiting planet, indicating that some systems with
large $\lambda$ of $30^{\circ}$ $\pm$ $21^{\circ}$ for TrES-1b
\citep{narita07}, and $62^{\circ}$ $\pm$ $25^{\circ}$ for HD 17156b
\citep{narita08}. Such planetary systems might have a chaotic
dynamical history that produce the present highly-inclined planets
\citep{chatt08}. Currently, it is hard to determine the proportion
of the exoplanet population with a significant inclination, which
requires precise measurements of $\lambda$ for a large quantity of
the systems. However, in this sense, it is extremely significant for
one to make a deeper understanding of complicated building process
for planetary formation, by investigating the configuration of
highly-inclined planets.

In this work, we extensively perform a series of groups of
simulations, to explore the planetary formation in the late stage
for the inclined system (application to OGLE-2006-BLG-109L, see
Table \ref{table1} for details), by considering major factors that
might have influence on final assembly of terrestrial planets. The
paper is structured as follows: \S 2  describes numerical setup for
investigation. \S 3 presents the results and the discussion is given
in section \S 4. Finally, we briefly summarize the main outcomes in
\S 5.

\section{NUMERICAL SETUP}

\begin{figure*}
\includegraphics[width=14cm]{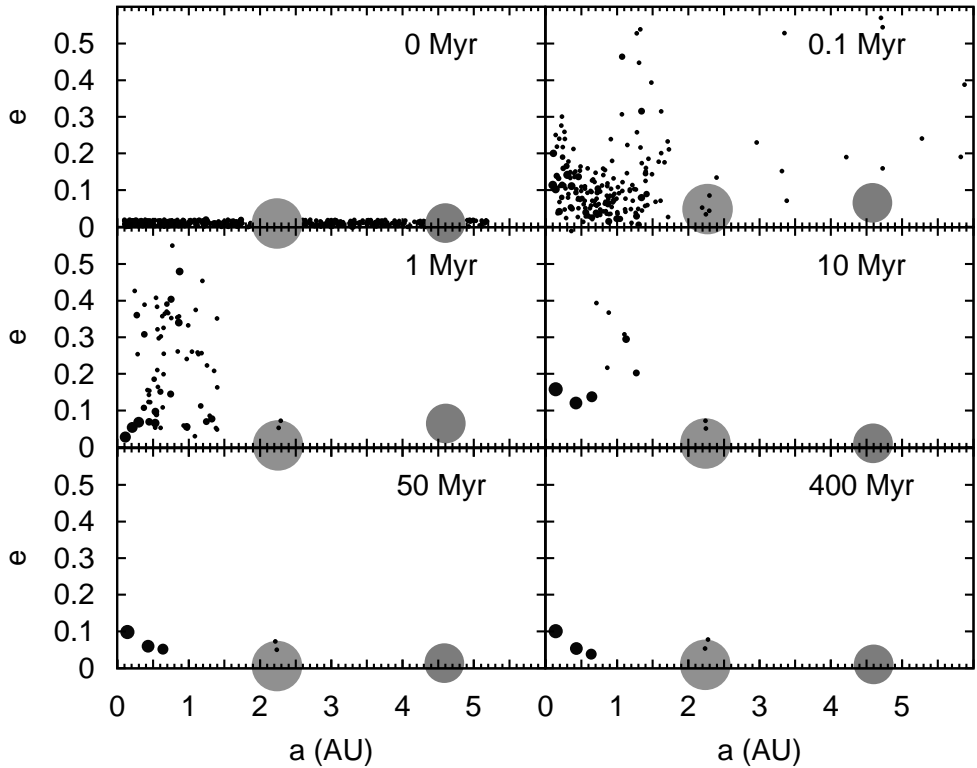}
\caption{Snapshots of a run in the simulation of Group 1. The panels
show the eccentricity versus semi-major axis for each surviving body
at $t = $ 0, 0.1, 1.0, 10, 50, and 400 Myr. The radii of
each object are related to their masses, with radius $\propto
m^{1/3}$. Two giants are, respectively, at 2.3 and 4.6 AU. Three
terrestrial planets have formed at $\sim 10$ Myr and remain stable
over $\sim 400$ Myr.} \label{fig1}
\end{figure*}

In our simulations, the initial conditions emulate the environment
at the beginning of the chaotic phase of terrestrial planet
formation \citep{cham01,raym04,raym06}, Moon-sized planetesimals and
Mars-sized planetary embryos are distributed in the disk, in company
with the OGLE-2006-BLG-109Lb,c analogues. The orbital elements  are
adopted from the observation data \citep{gaudi08}. For each run, we
assume the initial objects to follow a distribution with orbital
radius as $N\propto r^{-1/2}$, corresponding to the annular mass in
a disk with surface density \citep{weid77},
\begin{equation}
\Sigma\propto r^{-3/2}
\end{equation}
The Hill radius $R_H$ within which the gravity of planet dominates is,
\begin{equation}
  R_{H}=(M_{p}/3M_{*})^{1/3}a
\end{equation}
where $M_{p}$ and $M_{\odot}$ are the masses of planet and central
star respectively. Then the feeding zone of a planet at a given
position $a$  can be expressed as,
\begin{equation}
  F_{a}=C R_{H}
\end{equation}
where $C$ is a constant.
As the size of a planet grows, the corresponding feeding zone expands as well.
Nevertheless, we could obtain the isolation mass of planet at
a given orbital position by accumulating the total mass of
the materials in the feeding zone of the original disk,
\begin{equation}
  M_{\rm iso}=2\pi a \cdot 2F_{a} \cdot \Sigma_p  = 4 \pi a \cdot C \left(
  \frac{M_{\rm iso}}{3 M_*} \right)^{1/3} a \cdot \Sigma_p
\end{equation}
which gives,
\begin{equation}
  M_{\rm iso}\propto a^{2} \cdot M_{\rm iso}^{1/3} \cdot \Sigma_p \propto a^{3/4}
\end{equation}
Therefore, we set the masses of planetary embryos and planetesimals
following the relationship $M_{embryo}\propto a^{3/4}$, proportional
to the total mass in the embryo's feeding zone \citep{raym04}. On
the other hand, all the planetesimals in our simulations were set
with an equal mass of 0.017 $M_{\oplus}$.

\begin{table}
\centering \caption{Orbital parameters of two gas-giants
\citep{gaudi08}.} \label{table1}
\begin{tabular}{clclclc}
\hline
Planet & a (AU)  & e & Mass ($M_{Jup}$) \\
\hline
OGLE-2006-BLG-109L b  & 2.3  & 0.001  & 0.71\\
OGLE-2006-BLG-109L c  & 4.6  & 0.11  & 0.27\\
\hline
\end{tabular}
\end{table}

The initial orbital elements of each embryo and planetesimal were
randomly generated: argument of pericentre, longitude of the
ascending node, and mean anomaly of each small object were randomly
set between $0^{\circ}$ to $360^{\circ}$; the eccentricities range
from 0 to 0.02, while the inclination vary from $0^{\circ}$ to
$1^{\circ}$. For giant planets, to investigate terrestrial planetary
formation in an inclined configuration, the inclination of the outer
giant planet changes for each group but that of the inner gas-giant
remains, in the meantime all other orbital data keep unchanged in
the each initial run.

\begin{table}
\centering \caption{Simulation details of each group.}
\label{table2}
\begin{tabular}{clclclc}
\hline
Number & Embryos  & Planetesimals & Total Mass ($M_{\oplus}$) \\
\hline
Group 1  & 30  & 470  & 10\\
Group 2  & 30  & 470  & 10\\
Group 3  & 50  & 950  & 15\\
Group 4  & 30  & 470  & 10\\
Group 5  & 30  & 470  & 10\\
\hline
\end{tabular}
\end{table}

\begin{figure*}
\includegraphics[width=14cm]{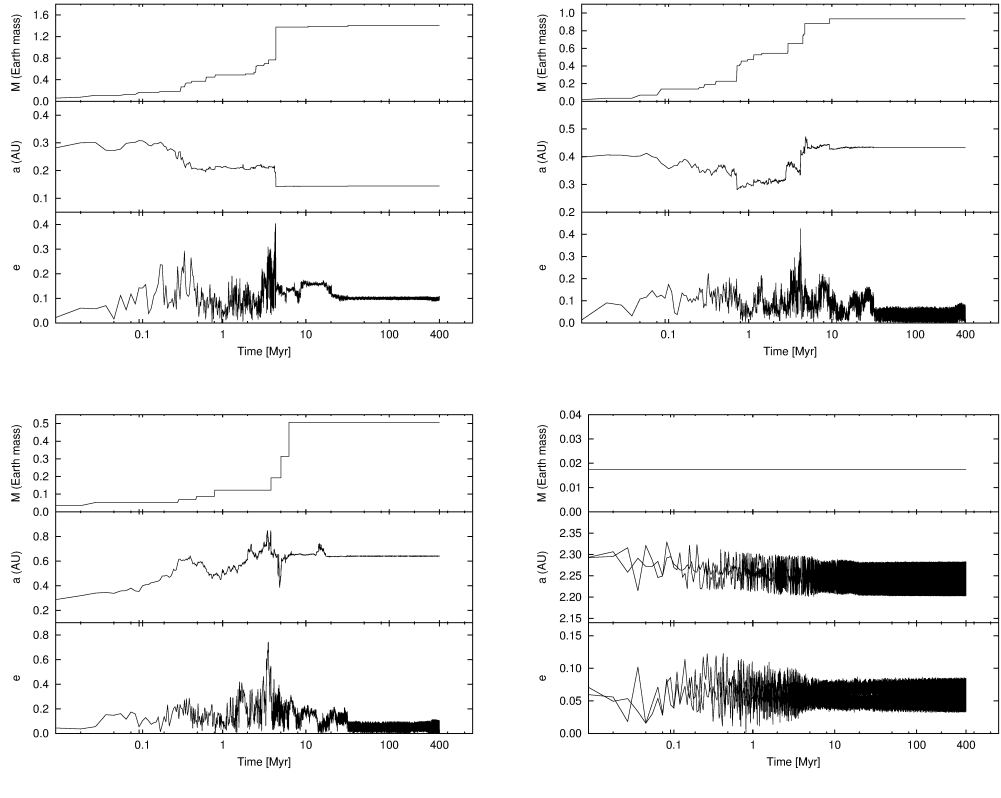}{h}
\caption{Mass, semi-major axis, and eccentricity versus time of all
the survivals in the run in Figure \ref{fig1}. Three terrestrial
planets formed at $\sim$ 10 Myr and remain in stable orbits
afterwards. Note that two planetesimals survive at the 1:1 MMR with
respect to OGLE-2006-BLG-109b, as shown in the right bottom panel.}
\label{fig2}
\end{figure*}

As mentioned, one of the main goals of this work is to explore
terrestrial planet formation in the late stage under the
circumstance of highly-inclined planetary system, where the mutual
inclination of the outer giant planet is taken into account.
Consequently, we carried out five groups of simulations, containing
46 runs in total. The major difference between each group is the
distribution range of the initial bodies. In all groups, two giant
planets were set to emulate the OGLE-06-109L system in each run,
with initial orbital parameters ($M_{P}$, $a$, $e_{p}$) $=$
(0.71\,$M_{jup}$, 2.3\,AU, 0.001) and (0.27\,$M_{jup}$, 4.6\,AU,
0.11)\citep{gaudi08}, as shown in Table \ref{table1}.  We adopt the
stellar mass as 0.5 $M_{\odot}$\citep{gaudi08}, and assume its
radius to be $\sim$ 0.8 $R_{\odot}$.  For each group, the simply
difference between individual runs is the inclination of the outer
giant, OGLE-2006-BLG-109Lc. Some details of each group are listed in
Table \ref{table2}, which is also summarized as follows,
 \begin{itemize}
   \item {\bf Group 1} - 5 runs. In each run, embryos and
     planetesimals were distributed in the range 0.1\,AU $< a <$ 5.2\,AU.
     The  inclination of OGLE-2006-BLG-109Lc in each run is adopted from $0^{\circ}$
     to $40^{\circ}$, in increments of $10^{\circ}$.
   \item {\bf Group 2} - 14 runs. In each run, embryos and
     planetesimals  were in the range 0.3\,AU $< a <$ 5.2\,AU.
     The  inclination of OGLE-2006-BLG-109Lc in each run is from $0^{\circ}$
     to $10^{\circ}$ in increments of $1^{\circ}$,
     and from $10^{\circ}$ to $40^{\circ}$ in increments of $10^{\circ}$.
   \item {\bf Group 3} - 17 runs.  In each run, embryos and
     planetesimals  were in the range 0.3\,AU $< a <$ 5.2\,AU.
     The  inclination of OGLE-2006-BLG-109Lc in each run is from $0^{\circ}$
     to $10^{\circ}$ in increments of $1^{\circ}$,
     and from $10^{\circ}$ to $40^{\circ}$ in increments of $5^{\circ}$.
  \item {\bf Group 4} - 5 runs. In each run, embryos and
    planetesimals  were distributed in the range 0.1\,AU $< a <$ 10\,AU.
     he inclination of the outer giants in each run is from $0^{\circ}$
     to $40^{\circ}$, in increments of $10^{\circ}$.
   \item {\bf Group 5} - 5 runs. In each run,  embryos and
     planetesimals  were distributed in the range 0.3\,AU $< a <$ 10\,AU.
     The inclination of the outer giants in each run is from $0^{\circ}$
     to $40^{\circ}$, in increments of $10^{\circ}$.
 \end{itemize}

Both embryos and planetesimals were considered to gravitationally
interact with each other in our simulations. Objects were allowed to
collide, and we assume that they merge into a single body with no
fragmentation after a collision. All 46 runs were performed using a
hybrid symplectic algorithm provided by the MERCURY integration
package \citep{chambers99}. Each run evolved for 400 Myr, with a
time step length no more than 3.0 days ($\sim$ a twentieth of a
period for the body at 0.3 AU). The Bulirsch-Stoer tolerance for all
runs is of $10^{-12}$. In most of the simulations, energy is
conserved to better than 1 part in $10^{-3}$, and the angular
momenta conserved in $10^{-11}$. Next, we will briefly introduce
simulation outcomes.

\section{RESULTS}

From the simulations, we can note that a typical accretion scenario
occurs in the late stage planet formation \citep{cham01, ji11}.  In
the beginning, the planetary embryos and planetesimals are quickly
excited to highly-eccentric orbits due to dramatic perturbations
arising from OGLE-2006-BLG-109Lb,c. Subsequently, frequent orbital
crossings emerge when the objects approach each other, which may
result in violate collisions amongst planetesimals and embryos.
Herein a large portion of the runs show that the planetesimal disk
becomes quite turbulent within 1 Myr, while in some cases their
chaotic period of movement could last a little longer, and most of
the initial objects were removed by ejections or collisions.

Fig.~\ref{fig1} illustrates that six snapshots of one run in Group
1. The figure shows that three terrestrial planets emerged at $\sim$
10 Myr and their orbits kept stable over subsequent evolution.
Moreover, at the end of this run, there are also two survivals
locked in the 1:1 Mean Motion Resonance (MMR) with
OGLE-2006-BLG-109Lb, which reminds us of the Trojan/Greek population
of asteroids at triangular points for Jupiter. In our simulations,
survived Trojans are not rare to find, for they are simply
unaccreted planetesimals.

The dynamical evolution process within 10 Myr for the bodies plays a
key role in creating members at final configurations. As shown in
Fig.~\ref{fig2}, it represents the mass, semi-major axis, and
eccentricity versus time for all final bodies. We observe that three
terrestrial planets had finally made to reach about Earth-size mass
in the stage. The eccentricities of each body change violently
during this chaotic stage, then they were damped down to $\leq$ 0.1.
In the final, the orbits of three terrestrial planets remained
stable afterwards. However, two planetesimals were trapped into 1:1
MMR with OGLE-2006-BLG-109b, and their orbital parameters are nearly
unchanged over the timescale of 400 Myr.

To summarize all outcomes, we note that the terrestrial planet
formation is very common in the simulation, as a great number of
runs may remain two or more planets eventually. In addition, over
2/3 of the runs have 1-3 survival planetesimals at the 1:1 MMR with
respect to OGLE-2006-BLG-109Lb. In the cases of the initial embryos
and planetesimals extending 10 AU, the planetary accretion in the
outer disk may still work. In such cases, a significant portion of
objects could stay at the region 7-15 AU in the end.

\subsection{Dependence on the outer highly-inclined giant}

\begin{table*}
\centering \caption{Statistic results for the simulations Group 3.
The columns are, respectively, the mutual inclination, the number of
ejected bodies, the objects collided with giant planets and central
star, the remaining mass (in Earth-mass) and the surviving mass
fraction for each run.} \label{table3}
\begin{tabular}{clclclclclcl}
\hline
Inclination ($^{\circ}$) &Ejection  &Giants &Star  &Remaining Mass ($M_{\oplus}$) &  Surviving Mass Fraction \\
\hline
0 &626  &69  &20 &3.404  &23\% \\
1 &669  &43  &26 &3.360 &22\% \\
2 &665  &54  &23 &3.450 &23\% \\
3 &664  &48  &20 &3.122 &21\% \\
4 &668  &37  &24 &3.526 &23\% \\
5 &686  &46  &28 &2.954 &20\% \\
6 &705  &49  &30 &2.701 &18\% \\
7 &707  &39  &28 &3.230 &21\% \\
8 &725  &40  &39 &2.121 &14\% \\
9 &690  &53  &37 &2.460 &16\% \\
10 &748  &43  &31 &1.954 &13\% \\
15 &743  &43  &53 &1.491 &10\% \\
20 &746  &47  &76 &0.6314 &4\% \\
25 &752  &50  &89 &0.6184 &4\% \\
30 &711  &57  &107 &0.1441 &1\% \\
35 &699  &61  &105 &0.02275 &0\% \\
40 &685  &63  &120 &0.01137 &0\% \\
\hline
\end{tabular}
\end{table*}

\begin{figure}
\includegraphics[width=9cm]{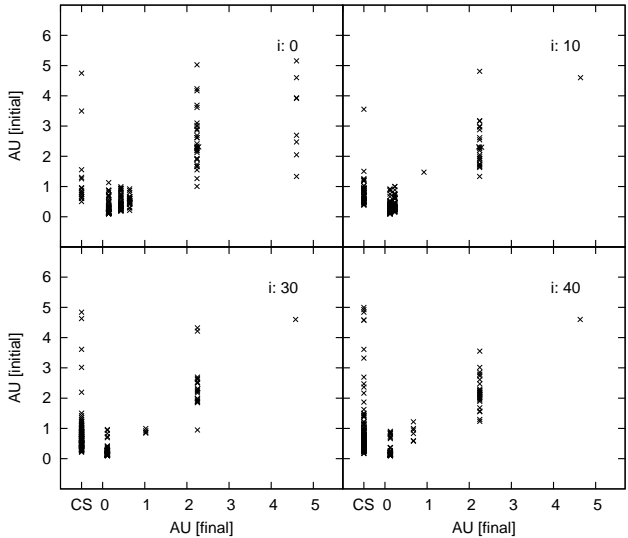}{h}\\
\caption{Trace of the destination of all objects from 4 runs in
Group 1. Each panel presents original versus final semi-major axes
for all objects -- except those scattered out of the system ($a >
100$ AU) -- of which the mutual inclination for the giants are,
respectively, $0^{\circ}$, $10^{\circ}$, $30^{\circ}$, and
$40^{\circ}$. The vertical axis indicates the original semi-major
axis of each body, while the horizontal axis indicates the final
value (CS denote the central star).} \label{fig3}
\end{figure}

\begin{figure}
\includegraphics[width=9cm]{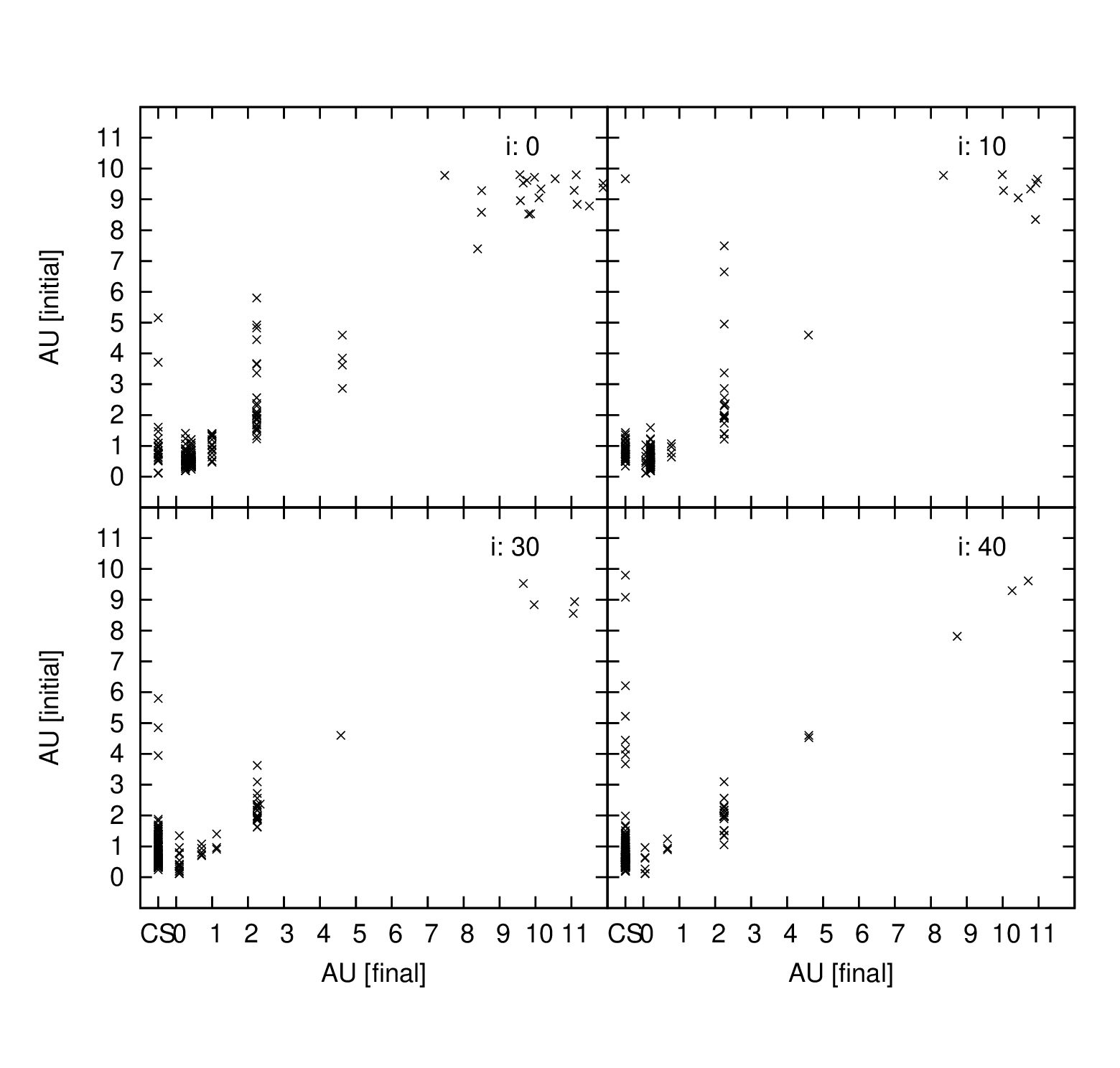}{h}\\
\caption{Trace of the destination of the objects from 4 runs in
Group 4 \textemdash layout is the same as for Figure \ref{fig3}. In
these runs, there are also some planetesimals survived in the outer
planetesimals disk.} \label{fig4}
\end{figure}

With an increasing mutual inclination between two giants, a great
many of smaller planetesimals either scattered out of the system, or
had been directly engulfed by the central star. Therefore a few
number of objects were left in the planetary disk, leading to the
difficult of making terrestrial planets. Table \ref{table3} presents
the statistic results over the simulations of Group 3, in which each
run has an initial population of 1000 bodies. The table summarizes
the number of scattered bodies, those hit into giants and host star,
the remaining mass and the accretion rate. For a moderate mutual
inclination ($i_m$) of $\sim 10^{\circ}$ to $25^{\circ}$, the
population of ejected bodies could be much more than that of $i_m <
5^{\circ}$. For a higher $i_m$ ($> 25^{\circ}$), the number of
ejected objects decreases. In this sense, a large relative
inclination of two giants seems to be less effective in throwing out
the embryos or planetesimals. However, on the other hand, in the
highly-inclined configurations the population that hit into the
central star could be significantly enhanced. As shown in Table
\ref{table3}, with $i_m < 10^{\circ}$, simply a small number of
planetesimals, say 20 - 40 objects, entered into the central star;
while for a larger mutual inclination, the number could rise nearly
linearly. Moreover, in the case of $i_m = 30^{\circ}, 40^{\circ}$,
the bodies that originally lie in the outer disk might run into the
central star over shorter timescale. In a word, the relatively
inclined orbital configuration do play a significant role in the
formation of terrestrial planets, inferring that the lower mutual
inclination of two giants, the more effective for planetary
accretion.

Fig.~\ref{fig3} shows the destination of the initial objects in
simulation of Group 1, except those were driven out of the systems.
The vertical axis indicates the original semi-major axis of each
body, while the horizontal line means the final. The figure shows
that along with the augment of the mutual inclination of two jovian
planets, the number of bodies that hit into the central star may
increase notably.  Especially, one may notice that the situation is
more apparent in the inner planetesimal disk at 0.1 - 1.5 AU.
Furthermore, Fig.~\ref{fig3} shows that in the case of $i_m =
30^{\circ}, 40^{\circ}$, a large proportion of the objects in the
inner disk could finally rush into the host star, leading to few
residuals in the disk for further planetary accretion. This trend is
in a good agreement with those as shown in Table \ref{table3} for
Group 3. Again, Fig.~\ref{fig4} shows the destination of  initial
objects in Group 4, which reveals a similar trend like
Fig.~\ref{fig3}. A large number of bodies that finally run into the
central star, may cause the deficiency in mass for the inner disk to
produce more terrestrial planets, and a scarcity of survivals in the
outer disk.

\begin{figure*}
\includegraphics[width=15cm]{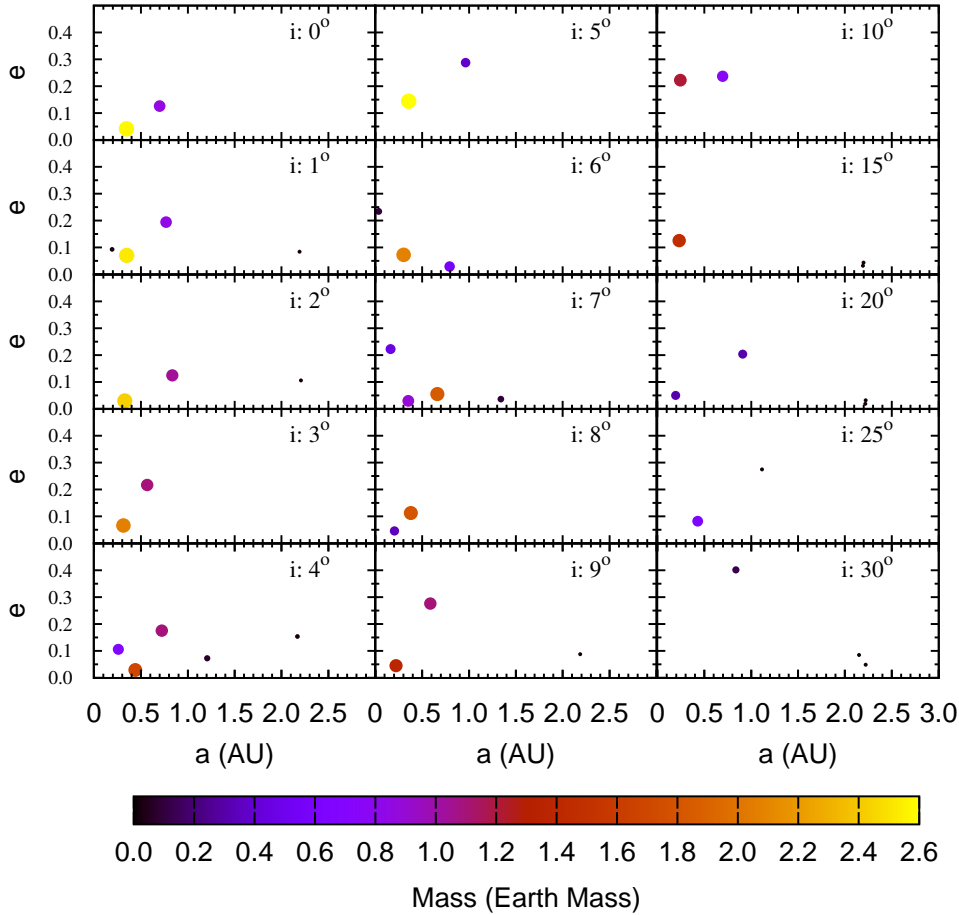}\\
\caption{All the final configurations of the runs that have
terrestrial planets formed in Group 3. The panels show the
eccentricity versus semi-major axis for each surviving body at the
end of each run. The radii and colors of each object are related to
their masses, with radius $\propto m^{1/3}$. The mutual inclination
of the two giants is labeled in each panel. Note that the mass of
formed planets decreases as the mutual inclination between the
giants increase.} \label{fig5}
\end{figure*}

As mentioned previously, a great many of bodies had been either
driven out of the system, or collided with the giant planets or
central star for inclined  configuration, leading to the difficulty
in planetary accretion. Moreover, Fig.~\ref{fig5} shows the
remaining bodies for each run from Group 3 over 400 Myr, except for
$i_m = 35^{\circ}, 40^{\circ}$ where no terrestrial planet formed.
The figure clearly reveals that the mass of created planets
decreases as the mutual inclination grows. As shown in
Fig.~\ref{fig5}, the $i_m =30^{\circ}$ run illustrates that a
terrestrial planet with mass of $\sim 0.1 M_{\oplus}$ was finally
formed at $\sim$ 0.8 AU, indicative of inefficiency for planetary
accretion. To sum up, simulation results demonstrate that
terrestrial planetary formation is extremely related to the mutual
inclined configuration of the giant planets in the system.

\subsection{Dependence on the orbital distribution of initial bodies}

\begin{table*}
\centering \caption{Statistics over 20 selected runs from four
different groups, considering the mutual inclinations. The name of
each run denotes the initial distribution range and the relative
inclination between two giants, e.g., 10-0.1-00 indicates a run for
embryos and planetesimals initially ranging from 0.1 to $\sim$ 10
AU, with a $0^{\circ}$ mutual inclination for two giant planets. In
the following columns, the table lists the final number of remaining
terrestrial planets and planetesimals (referred as TP and PL,
respectively), the total remaining mass (in Earth-mass) and the
surviving mass fraction of formed terrestrial planets.}
\label{table4}
\begin{tabular}{clclclcccc}
\hline
Simulation  &Initial  &Final(TP + PL) &Remaining Mass ($M_{\oplus}$) &Surviving Mass Fraction (TP) \\
\hline
10-0.1-00  &500  &3+23 &3.090 &27\% \\
10-0.3-00  &500  &2+13 &2.832 &27\% \\
5.2-0.1-00  &500  &3+2  &2.852 &28\% \\
5.2-0.3-00  &500  &2+2 &1.900 &19\% \\
\hline
10-0.1-10  &500   &3+12  &2.630 &24\% \\
10-0.3-10  &500   &2+10 &1.585 &13\% \\
5.2-0.1-10  &500  &3+2  &2.469 &24\% \\
5.2-0.3-10  &500  &2+2   &1.604 &16\% \\
\hline
10-0.1-20  &500   &3+9 &1.946 &18\% \\
10-0.3-20  &500   &1+6  &0.777 &7\% \\
5.2-0.1-20  &500  &2+2  &1.864 &18\% \\
5.2-0.3-20  &500  &3+1   &0.931 &9\% \\
\hline
10-0.1-30  &500   &3+8 &1.174 &11\% \\
10-0.3-30  &500   &0+8  &0.381 &3\% \\
5.2-0.1-30  &500  &2+1  &1.334 &13\% \\
5.2-0.3-30  &500  &1+1  &0.210 &2\% \\
\hline
10-0.1-40  &500   &2+4 &0.436 &4\% \\
10-0.3-40  &500  &0+6 &0.222 &1\% \\
5.2-0.1-40  &500 &2+0  &1.363 &14\% \\
5.2-0.3-40  &500 &1+3   &0.229 &3\% \\
\hline
\end{tabular}
\end{table*}

\begin{figure*}
\includegraphics[width=15cm]{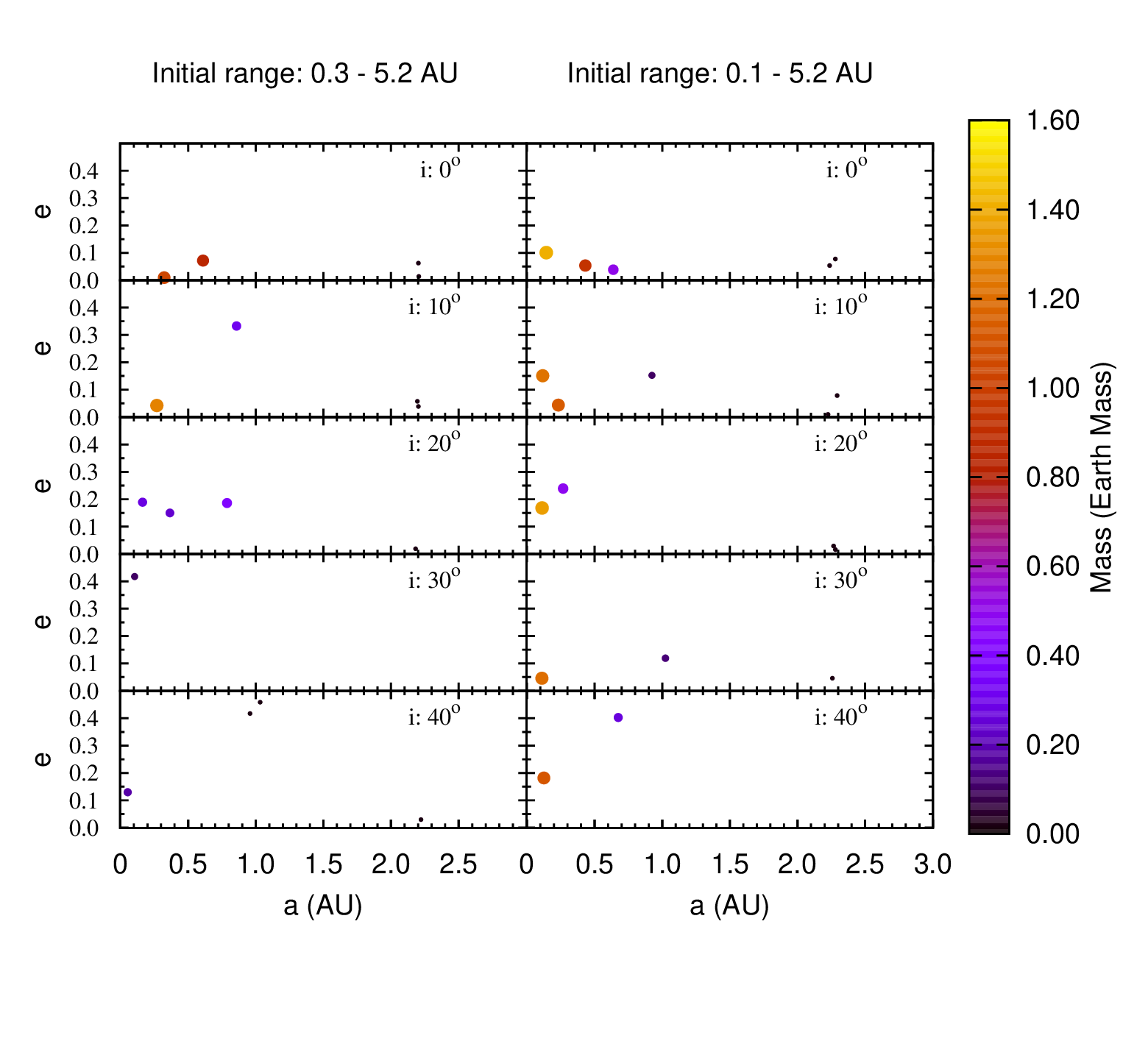}\\
\caption{The comparison of final configurations of 5 runs in Group 1
(where 0.1 AU $<a<$ 5.2 AU for initial objects) and Group 2 (0.3 AU
$<a<$ 5.2 AU), and the layout is the same as Figure \ref{fig5}. Note
that in the case of Group 1 more terrestrial planets could form.}
\label{fig6}
\end{figure*}

The different distribution ranges of initial bodies also play a
major part in yielding final members in the system for each
simulation. Table \ref{table4} introduces five contrast groups, and
in each run the number of the initial bodies is 500, with a total
mass of 10 $M_{\oplus}$, but they are distributed in various initial
orbits. As shown in Table \ref{table4}, in each group -- in which
all runs have equal mutual inclination relative to giant planets --
runs with an initial disk of an inner edge at 0.1 AU seem to have
advantage of occupying more survivals of terrestrial planets and
planetesimals, with a larger total mass of remaining bodies and
efficient accretion rate of the terrestrial planets.

Fig.~\ref{fig6} shows the outcomes of ten runs from Group 1 and 2,
to compare the overall results for different initial distribution
ranges. Similar to Table \ref{table4}, both the number and mass of
formed terrestrial planets in Group 1 (0.1\,AU $< a <$ 5.2\,AU) is
larger than those in Group 2 (0.3\,AU $< a <$ 5.2\,AU). Considering
the smaller semi-major axis of two giants in this system, planetary
embryos and planetesimals in the outer region of the planetesimals
disk would suffer severe gravitational influence from two giant
planets; hence, the planetesimals, locating at 1 $\sim$ 7 AU could
not remain stable under dramatic perturbation from them. The only
exception is the 1:1 MMR region with respect to the
OGLE-2006-BLG-109b, while no such Trojan was found for the outer
giant planet. This may explain an almost empty disk for all final
configurations, ranging from 1 $\sim$ 7 AU, in our simulations. On
the contrary, the inner part of the disk is strongly favor of
producing terrestrial planets, because of the relatively higher
density of the population and stronger influence of the central
star. In this sense, these cause the inner region of the planetary
disk, from 0.1 to 0.3 AU, to be very important for planetary
accretion for terrestrial planets.

\begin{figure*}
\includegraphics[width=14cm]{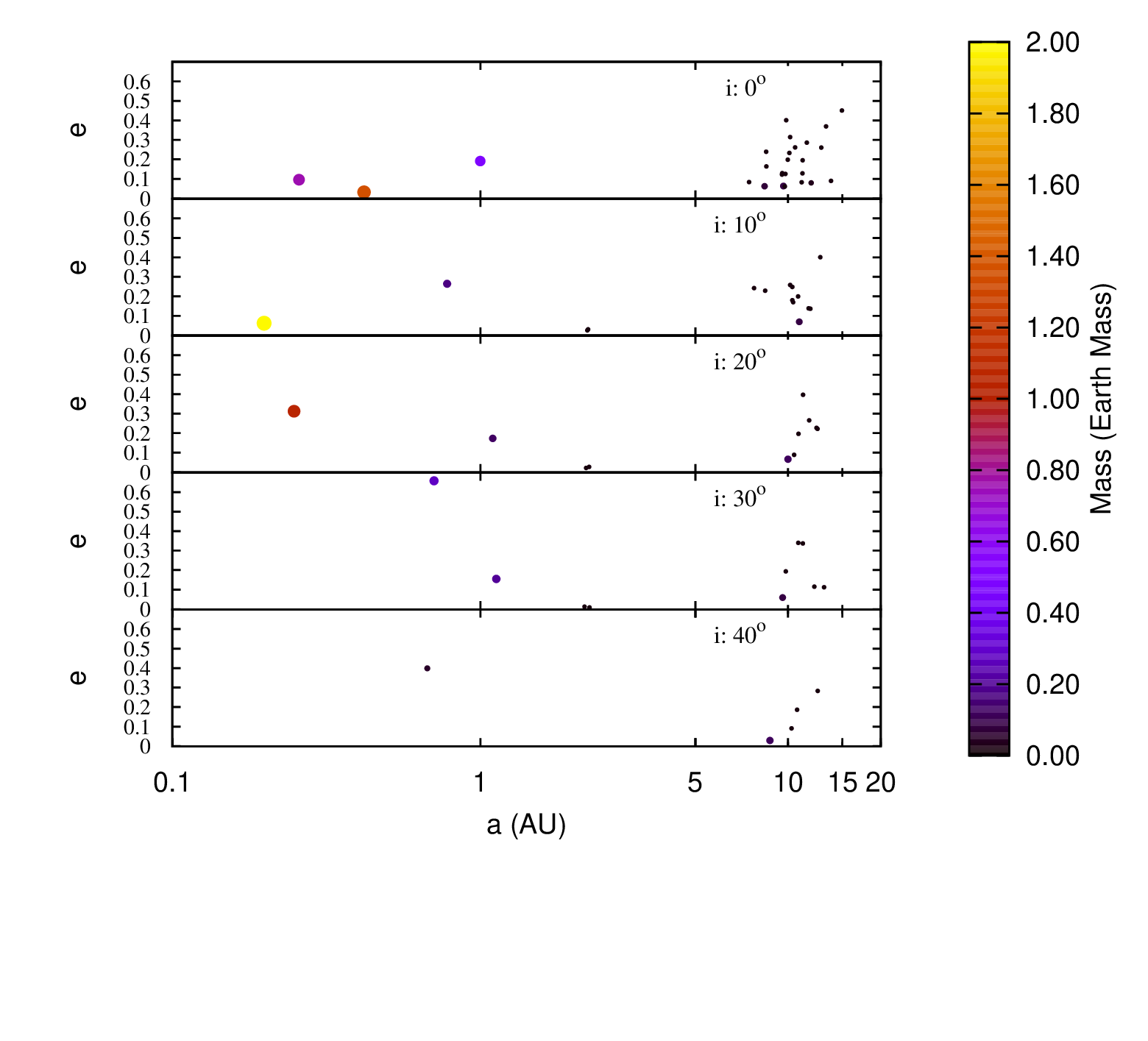}\\
\caption{Final configurations of 5 runs in Group 4. The layout is
the same as Figure \ref{fig5}, but the horizontal line is in a
logarithmic scale. The figure shows the terrestrial planetary
accretion becomes difficult in the case of the increasing mutual
inclination of the gas-giants. In the final, there exist several
survivals in the outer region of the disk extending to 7 AU.}
\label{fig7}
\end{figure*}

\begin{figure*}
\includegraphics[width=14cm]{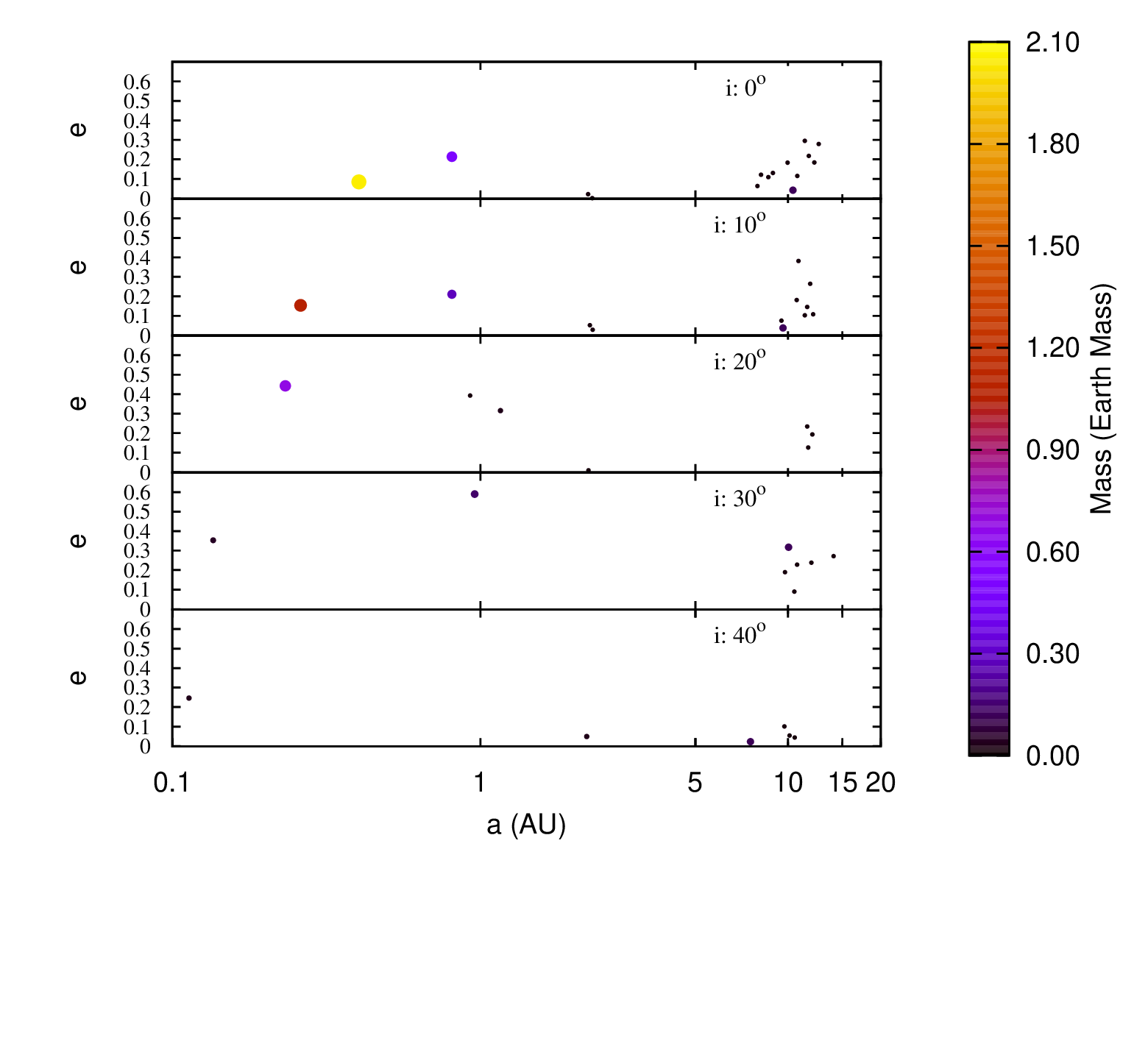}\\
\caption{All the final configurations of 5 runs in Group 4, the
layout is the same as Figure \ref{fig7}. In Comparison with Figure
\ref{fig7}, the runs of 0.1 $\sim$ 10 AU configuration would
relatively have less and smaller final planets than in the 0.3
$\sim$ 10 AU distribution cases.} \label{fig8}
\end{figure*}

The above feature is also reported in much wider distribution cases,
where the embryos and planetesimals originally located at an outer
edge of disk, extending to 10 AU. Figure \ref{fig7} shows the
terrestrial planetary accretion becomes more difficult when the
mutual inclination of the gas-giants increases. One may note that in
the final evolution there exist several survivals in the outer
region of the disk up to 7 AU.  In Figure \ref{fig8}, the runs of
0.1 $\sim$ 10 AU configuration would relatively have less and
smaller final planets than those in the 0.3 $\sim$ 10 AU cases.

In addition, in the cases of initial distribution up to 10 AU, there
also exist several survivals in the outer realm of the planetary
disk at 7 AU. Some survivals experienced several accretion, at a
limited rate, in comparison with those in the inner disk; others,
like Mar-sized protoplanets, emerged around the region of 7.5 $\sim$
11 AU in the end of simulations. Compared with our solar system,
these formed bodies may be in a rescaled position of Uranus, Neptune
and the Kuiper belt objects, implying that Neptunian-mass planet may
be formed if more bodies were initial placed there and abundant mass
were considered over much longer timescale evolution.

\subsection{Planetary formation in the habitable zone}

As mentioned, the OGLE-2006-BLG-109L was the first double planet
system, discovered by gravitational microlensing method
\citep{gaudi08}. Two companions occupy the masses of $\sim0.71$ and
$\sim0.27 M_J$, respectively, and each orbits its central star at
about 2.3 and 4.6 AU. The mass of the star is $\sim$0.5 $M_{\odot}$,
and this system is resemblance to a rescaled solar system, for the
mass ratio and the separation between them is comparable to those of
Jupiter and Saturn. The habitable zone around a star is defined as
the region where a terrestrial planet with $N_{2}$-$CO_{2}$-$H_{2}O$
atmosphere could maintain liquid oceans on its surface
\citep{kasting93}. As for this system, the HZ is about
0.25$\sim$0.36 AU from the star \citep{miga09}. Since the
OGLE-2006-BLG-109L planetary system bears very close similarity to
solar system, one may wonder to know whether it is possible to
harbor other planets, especially for potential Earth-like planets in
the HZ.  Recently, \citet{malhotra08} analyzed secular dynamics for
OGLE-2006-BLG-109L and presented a stable model to explain the
existence of two supposed terrestrial planets in the system.
\citet{miga09} and \citet{wang09} further showed that Earth-size
bodies might be formed in the HZ of the system.

\begin{figure*}
\includegraphics[width=14cm]{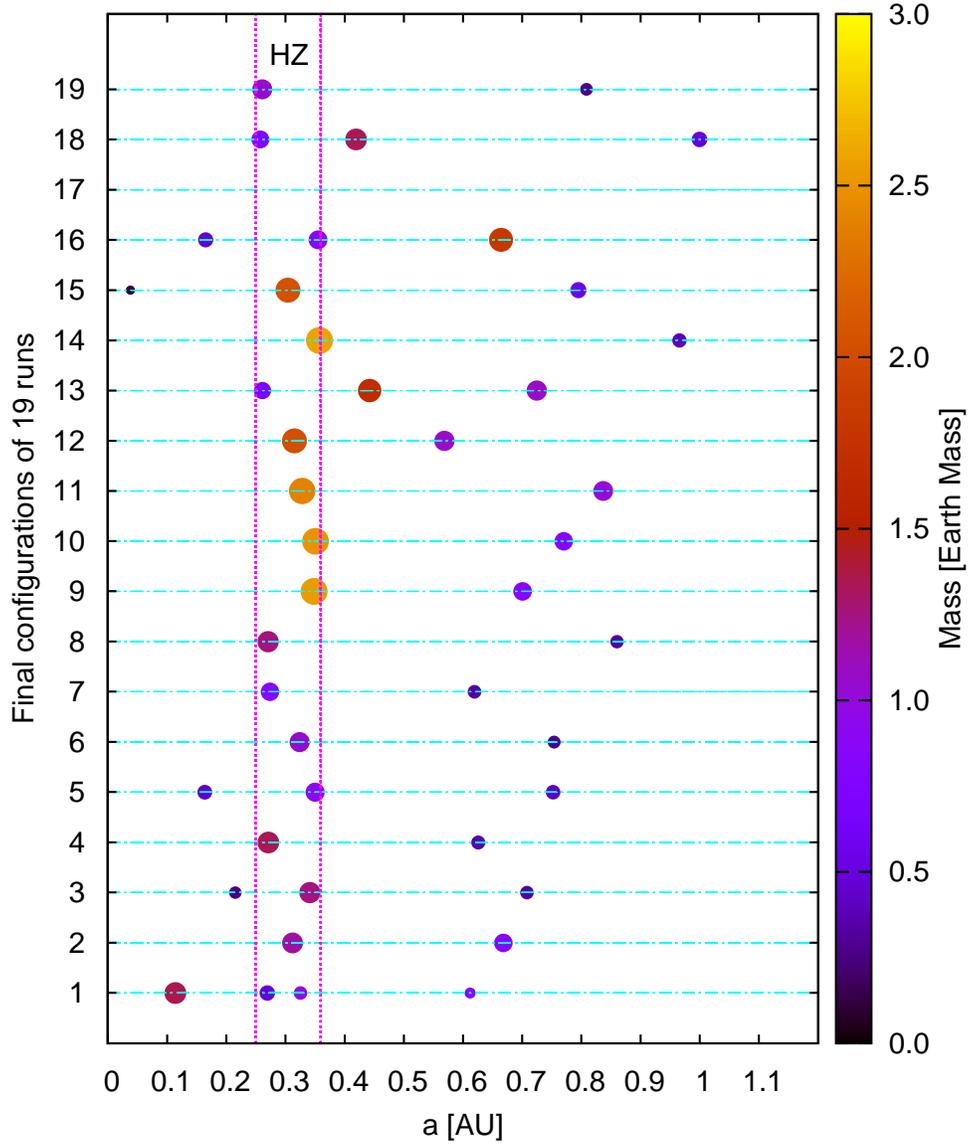}\\
\caption{Final configurations of 19 runs that have formed habitable
planets. The vertical axis indicates the number of each run. The
radii and colors of each object are related to their masses, with
radius $\propto m^{1/3}$. Two vertical dotted lines represents the
inner and outer boundaries of the HZ for this system. The figure
also shows that in the final runs for two terrestrial planets, one
resides in the HZ and the other locates in outer regime ranging from
0.4 to 1.0 AU.} \label{fig9}
\end{figure*}

From our simulations, we note that more than 40\% (19 out of 46) of
the total runs have formed planets in the HZ, within 0.25 $\sim$
0.36 AU \citep{miga09}. In fact, we find that in certain
configurations with a smaller mutual $i$ ($\leq 10^{\circ}$) may
have a great probability, e.g., 16 out of 22 runs formed a
terrestrial planet in the HZ. Fig.~\ref{fig9} shows all final
configurations of 19 runs have produced habitable planets at the end
of numerical simulation, where Run 1 $\sim$ 8 are adopted from Group
2, Run 9 $\sim$ 16 from Group 3, and  Run 17, 18 and 19 from Group
1, 4 and 5, respectively. Obviously, as illustrated in
Fig.~\ref{fig9},  most of the runs are from Group 2 and 3, which
have an initial distribution for embryos and planetesimals ranging
from 0.3 - 5.2 AU.

The common feature in these runs is indicative of the final
configuration to consist of two terrestrial planets, with one in the
HZ and the other in outer region from 0.4 to 1.0 AU. Typically, the
planet in the HZ seems to be more massive ($0.75 \sim 2.6
M_{\oplus}$), while the outer body has a smaller mass ($0.27 \sim
1.81 M_{\oplus}$). In addition, there also exists some
configurations (totally 6 runs) that three terrestrial planets
formed. In these cases, the other two planets do not reside in
habitable zone. However, no configuration of a single habitable
planet has been found in the final simulations, although it was
considered to be a stable model under the combinations of a low
eccentric orbit and some proper orbital parameters of jovian planets
\citep{miga09}.

The uniform characteristics of the formation of habitable
terrestrial planets in our simulations over 400 Myr, could be served
as a significant evidence of the stability of the HZ in the
OGLE-2006-BLG-109L system.

\section{DISCUSSION}

The OGLE-2006-BLG-109L system has drawn many researcher's attention
as it is resembling to a rescaled solar system. \citet{wang09}
showed that the stability of the region is within $a \leq 1.5$ AU
and $a \geq 9.7$ in their numerical simulations, and also pointed
out the HZ is broader enough for a stable Earth-mass planet. From
the runs, we find that terrestrial planets could emerge in the
regions where $0.1 < a < 1.13$ AU or $a
> 7.5$ AU. For $1.13 < a < 1.5$ AU, there may remain stable
for an additional planet as suggested by \citet{wang09}, where a
great many of initial bodies are ejected due to strong stirring of
two jovian planets, and they either scatter out of the system or
directly run into the central star.  As a result, simply a few
number of objects were left as accretion stuff for further planetary
formation, consequently none of the planets has finally formed in
this region. Herein we also confirm the stability of the HZ in this
system: 19 out of total 46 runs have a final planet locating in 0.25
to 0.36 AU, which is in good agreement with theirs.

\citet{malhotra08} found that secular resonance arising from the
massive planets in the system could drive an Earth-mass planet out
of the HZ, and an additional inner planet, with mass larger than
$0.3 M_{\oplus}$ at $a\leq 0.1$ AU, is required to maintain the
habitability of the OGLE-2006-BLG-109L system. Therefore, they
suggest a possible configuration with at least two additional
planets in order to support a potentially habitable Earth-like
planet in this system. However, in the 19 simulations that formed a
habitable planet, we show that most of final configurations bear one
larger planet ($0.75\sim2.6 M_{\oplus}$) in the HZ, and an
additional one or two less massive ($0.27\sim1.81 M_{\oplus}$)
planets in the outer region ranging from 0.4 to 1.0 AU. Simply one
run in our work may have a similarity to the configuration as
mentioned by \citet{malhotra08}, where three planets with parameters
($M_{P}$, $a$) $=$ (0.1\,$M_{\oplus}$, 0.0385\,AU),
(2.05\,$M_{\oplus}$, 0.3044\,AU), and
(0.55\,$M_{\oplus}$,0.7951\,AU). Our numerical results a bit differ
from their theoretical analysis in the initial distribution of inner
boundary (where we set to be 0.1 AU) for planetesimals and embryos.
However, from viewpoint of habitability, our simulations are in
consistent with their suggestion that at least two additional
planets are required to support a potentially habitable Earth-like
planet in this system.

In the simulations, the configurations with a single stable
habitable planet as mentioned by \citet{miga09} do not occur, where
a terrestrial planet could survive to move on low-eccentric orbit
for the combination of the apsidal angle of jovian orbits,
eccentricities and semi-major axes consistent with the observations.
However, our work differs from those in that in our simulations we
adopt the highly-inclined configuration between two jovian planets,
and we also have set a much larger total mass and more initial
bodies than those of \citet{miga09}, where they simply considered
less bodies - 50 Moon-sized planetary embryos between 0.2 - 2 AU,
leading to a great difficulty in producing more terrestrial planets.
Nonetheless, our work is in concurrence with their study in the fact
that most of the runs formed a habitable planet have an additional
body with a similar mass in the region 0.6 $\sim$ 0.8 AU.

Recently, \citet{wang10} investigated the formation of terrestrial
planets in the early stage of planet formation at the time the
gaseous disk has not dissipated yet. They indicated that the
formation and the final mass of the habitable planet in the
OGLE-06-10L system depends on the stellar accretion rate $\dot{M}$,
which means the rate at which the star accretes gas from the disk,
and the speed of Type I migration, which is more significant. In
this work, we take into account formation of terrestrial planets in
the late stage after the gas disk has dissipated. We concentrate on
the various distribution range of the initial bodies and mutual
inclination between two gas giants. Generally, the higher relative
inclination, the harder the planetary accretion. This is because the
relative highly-inclined outer planet may disturb the embryos and
planetesimals  in a more effective manner, which results in a large
number of objects thrown out of the system or run into the central
star, and then a limited number of bodies are left in the disk for
further planetary accretion.

With a relatively higher inclination, a large amount of the bodies
would be excited to be swallowed by the central star, especially
those in the inner region 0.1 to 1.5 AU. Moreover, the total mass of
the scattered bodies also increases as the mutual inclination does.
In a word,  the above-mentioned two factors do no good to form
terrestrial planets for the highly-inclined configurations.

Given two giants in this system with close-in orbits, the embryos
and planetesimals in the inner disk were under strong gravitational
influence arising from the jovian planets. Our results show that the
total accretion mass in the runs of the initial distribution at 0.1
AU is much larger than those distributed at 0.3 AU. In this sense,
it may imply that the location of the inner planetesimal disk, as
well as the number of the initial bodies, plays an important part in
the planetary accretion. On the other hand, embryos and
planetesimals between 1 $\sim$ 7 AU could not survive so long owing
to dramatic stirring of two gas-giants, except those at 1:1 MMR
location with the inner planet, which reminds us Trojan/Greek Group
of Jupiter. In two groups, the accretion does still work in the
outer region beyond 7 AU, with a moderate rate and limited speed.
Figs.~\ref{fig7} and \ref{fig8} show that several Mar-sized
protoplanets emerge between 7.5 to 11 AU over 400 Myr. In addition,
there are also many survival planetesimals at 7 $\sim$ 16 AU.
Suppose a larger number and a longer integration time were chosen at
the beginning of these runs, it may assume that Neptune-like planets
might appear in the end due to further planetary accretion in the
outer disk, similar to the outer structure of our solar system.

\section{SUMMARY}

In this work, we have carried out several groups of simulations,
with 46 runs in total, to investigate the planet formation in the
late stage for the OGLE-2006-BLG-109L system. We may summarize the
main results as follows:

Firstly, the simulation outcomes show that it is quite common for
terrestrial planet or possible habitable planets to finally form
residing in 0.25 $\sim$ 0.36 AU. In addition, our work may imply
that the OGLE-2006-BLG-109L system bears a great probability of
harboring one or additional terrestrial planets.

Secondly, the comparison of results for different groups suggests
that the efficiency of planetary accretion may decrease as the
relative inclination of two giant planets increases.

Finally, our results further show that embryos and planetesimals in
the 0.1 - 0.3 AU may play an important role in planetary accretion.
Moreover, we find that the bodies initially locate in the $\sim$ 1.1
AU - 7 AU of planetary disk are mostly unstable over the dynamical
evolution, which leads to no survivals for embryo or planetesimal at
this range in all simulations, except those at $\sim$ 2.2 AU close
to 1:1 MMR with OGLE-2006-BLG-109Lb. On the other hand, the
planetary accretion could continue to happen but keep a much slower
pace in the outer planetary disk for those runs with initial
semi-major axes extending 10 AU, where a significant number of
residuals eventually remain in the system in the simulations.

Currently, the Kepler mission is surveying 156,000 faint stars for
transiting planets as small as Earth, e.g., Kepler-9
\citep{holman10} and Kepler-11 \citep{liss11}. Our results here
imply the great probability to occupy stable terrestrial planets in
the OGLE-2006-BLG-109L system, which should be carefully examined by
forthcoming observations. With the help of high precision of
ground-based and spaceborne measurements on the exoplanets, we might
expect the discovery of solar system analogues in the near future.

\section*{Acknowledgments}
We thank our referee, J.E. Chambers, for a prompt report and good
comments that helped to improve the contents. We are grateful to the
support by the National Natural Science Foundation of China (Grants
10973044, 10833001), the Natural Science Foundation of Jiangsu
Province, and the Foundation of Minor Planets of Purple Mountain
Observatory.

\label{lastpage}

\end{document}